\documentclass[letterpaper, 10 pt, conference]{ieeeconf}  

\IEEEoverridecommandlockouts                              

\usepackage{amsmath} 
\usepackage{amssymb}  
\usepackage[font=footnotesize]{caption}
\usepackage{subfigure}

\usepackage{mathrsfs}
\DeclareMathAlphabet{\mathpzc}{OT1}{pzc}{m}{it}

\usepackage[noadjust]{cite}

\usepackage{tikz}
\usepackage{xspace}

\DeclareRobustCommand{\legendline}[1]{\hspace{-3pt}
\tikz[#1,line width=0.4mm,baseline=-0.5ex]{\draw (0,0) -- (.35,0);}
\hspace{-3pt}}
\DeclareRobustCommand{\legendsquare}[1]{\hspace{-2pt}
  \tikz[baseline=-0.5ex]{\node[#1, inner sep=.8ex, outer sep=0] (a) {};}
\hspace{-2pt}}
\definecolor{mblue}{rgb}{0,0.4470,0.7410}
\definecolor{morange}{rgb}{0.8500,0.3250,0.0980}
\definecolor{myellow}{rgb}{0.9290,0.6940,0.1250}
\definecolor{mpurple}{rgb}{0.4940,0.1840,0.5560}
\definecolor{mgreen}{rgb}{0.4660,0.6740,0.1880}
\definecolor{mcyan}{rgb}{0.3010,0.7450,0.9330}
\definecolor{mred}{rgb}{0.6350,0.0780,0.1840}
\definecolor{mgreenblue}{rgb}{0.0,1.0,0.5}
\definecolor{parulablue}{rgb}{0.2431,0.1490,0.6588}
\definecolor{parulalblue}{RGB}{39,151,235}
\definecolor{parulagreen}{RGB}{129,204,89}
\definecolor{parulayellow}{RGB}{249,251,21}

\newcommand{\norm}[1]{\left\lVert#1\right\rVert}

\newcommand{\ltwo}{\ensuremath{\mathcal{L}_2}\xspace}

\newcommand{\m}[1]{\mathcal{#1}}
\newcommand{\mr}[1]{\mathrm{#1}}

\newcommand{\PK}[1]{#1}
\newcommand{\col}{\mathrm{col}}

\title{\LARGE \bf
Pitfalls of Guaranteeing Asymptotic Stability in LPV Control of Nonlinear Systems
}

\author{P.J.W. Koelewijn, G. Sales Mazzoccante, R. T\'oth and S. Weiland
\thanks{This work has received funding from the European Research Council (ERC) under the European Union’s Horizon 2020 research and innovation programme (grant agreement nr. 714663).}
\thanks{P.J.W. Koelewijn, G. Sales Mazzoccante, R. T\'oth and S. Weiland are with the Control System Group, Faculty of  Electrical Engineering, Eindhoven University of Technology, 5600 MB Eindhoven, The Netherlands
        {\tt\small $\lbrace$p.j.w.koelewijn, r.toth, s.weiland$\rbrace$@tue.nl}.}%
}

\begin{document}

\maketitle
\thispagestyle{empty}
\pagestyle{empty}

\begin{abstract}
Recently, a number of counter examples have surfaced where Linear Parameter-Varying (LPV) control synthesis applied to achieve asymptotic output tracking and disturbance rejection for a nonlinear system, fails to achieve the desired asymptotic tracking and rejection behavior even when the scheduling variations remain in the bounded region considered during design. It has been observed that the controlled system may exhibit an oscillatory motion around the equilibrium point in the presence of a bounded constant input disturbance even if integral action is present. This work aims at investigating how and why the baseline Lyapunov stability notion, currently widely used in the LPV framework, fails to guarantee the desired system behavior. Specifically, it is shown why the quadratic Lyapunov concept is insufficient to always guarantee asymptotic stability under reference tracking and disturbance rejection scenarios, and why an equilibrium independent stability notion is required for LPV stability analysis and synthesis of controllers. The introduced concepts and the apparent pitfalls are demonstrated via a simulation example.
\end{abstract}

\vspace{-.3em}
\section{Introduction}\label{sec:intro}
\begin{figure}
	\centering
	\subfigure[NL system.]{\hspace{.5em}
	\includegraphics[scale=1]{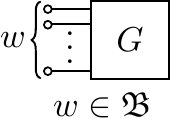}
	\hspace{.5em}
	}
	\subfigure[LPV embedding.]{\hspace{.5em}
	\includegraphics[scale=1]{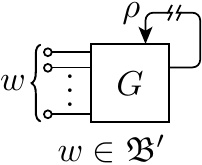}
	\hspace{.5em}
	}
	\hspace{.5em}
	\subfigure[Relation resulting behavior.]{
	\includegraphics[scale=1]{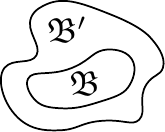}
	\hspace{.5em}
	}
	\vspace{-.5em}
	\caption{LPV embedding of an NL system and the resulting behaviors: $\mathfrak{B}$: solution set of $G$; and $\mathfrak{B}'$: solution set of the LPV model over $\rho\in \mathbb{P}$.}
	\label{fig:lpvemb}
	\vspace{-2em}
\end{figure}

\vspace{-.3em}
The ever-growing performance demands of today's industry, have resulted in increased system complexity requiring tools beyond the \emph{Linear Time-Invariant} (LTI) framework.
As a consequence, several \emph{Nonlinear} (NL) modeling and control methods have been developed \cite{Isidori1995,Nijmeijer2016,Khalil2015}. One of the drawbacks of these NL methods is that they often lack the systematic controller design procedures and performance shaping approaches of the LTI framework. As an alternative, varying concepts using linear proxy models have appeared and have extended the systematic analysis and synthesis tools of the LTI framework. Among these, the \emph{Linear Parameter-Varying} (LPV) framework has become a popular approach \cite{Toth2010}.
LPV models are capable of describing NL behavior in terms of a linear dynamical relation whose mathematical  description depends on a measurable, time-varying parameter, the so-called \emph{scheduling-variable} $\rho$ that resides in an a priori known/assumed set $\mathbb{P}$. 
Besides of early work on \emph{gain-scheduling} and \emph{local synthesis} methods \cite{Rugh2000}, the main interest towards the LPV framework originates from the observation made in the end of the 90's that it is possible to fully embed the solution set, i.e. behavior $\mathfrak{B}$, of an NL system (see Fig.~\ref{fig:lpvemb}.a) into the solution set $\mathfrak{B}^\prime$ of an LPV representation. This LPV \emph{embedding} of $\mathfrak{B}$ is achieved by extracting $\rho$ as a latent variable (see Fig.\ \ref{fig:lpvemb}.b) such that if the loop is disconnected, then the ``remaining" signal relations are linear. In this process, $\rho$ becomes a function of the output, input, and/or state of the original system representation through a so-called \emph{scheduling map} $\mu$. 
Assuming that $\rho$ is \emph{independent} of {the} output, input, and/or state signals \cite{Toth2010} provides a linear varying representation of the NL system, but this conceptual disconnection of $\rho$ introduces conservatism. This means the trajectories of the reformulated system form a behavior $\mathfrak{B}^\prime$ which contains $\mathfrak{B}$ (see Fig. \ref{fig:lpvemb}.c). {Hence,} any controller that realizes a desired operation on $\mathfrak{B}'$ for all possible trajectories of $\rho$, will also achieve the same objective if applied on $\mathfrak{B}$.  Through the obtained LPV representation, linearity can be exploited, resulting in convex optimization tools, which allow to ensure global stability, performance analysis and control synthesis with reduced computational complexity and more robustness when compared to other NL methods.  Due to these useful properties many powerful LPV analysis and control synthesis methods have appeared and have been applied to a wide range of industrial applications, see \cite{Hoffmann2015,Mohammadpour2012} and the references therein for more details.

Recently, in \cite{Scorletti2015,Koelewijn2019a}, it has been shown by counterexamples that \PK{the notion of \ltwo-gain stability is not sufficient to guarantee asymptotic output tracking and disturbance rejection for NL systems using LPV control methods}. In simulation studies, it has been shown that the controlled system can exhibit oscillatory motion around an equilibrium point, defined by a reference signal, in the presence of a bounded constant input disturbance, even if integral action is present in the control loop. In fact, such a problem may occur with other linear proxy model based frameworks building on the extension of the LTI framework.
Despite of the remedies that have been proposed in \cite{Scorletti2015,Koelewijn2019a}, no further analysis has been given why using $\ltwo$-gain performance and stability the LPV controllers fail to guarantee \PK{expected} stability and performance requirements for NL systems, while for LTI systems no such problems exist. The main contribution of this work is providing an analysis of this question from a nonlinear (Lyapunov) stability point of view.
It is shown that the necessary conditions for asymptotic stability guarantees for LPV representations with scheduling signals dependent on the output/input or state signals associated with the NL system do not ensure the same guarantees for the represented NL system for equilibrium points other than zero. Thus, naively using the usual $\ltwo$-gain LPV control methods to ensure stability and performance for an NL system \PK{for} reference tracking and\PK{/or} disturbance rejection could result in unexpected \PK{performance of the closed-loop system}. 

{The paper is structured as follows. In Section \ref{sec:problem}, the problem setting is described and an example is given illustrating the problem. Section \ref{sec:lyap} describes the current stability analysis of LPV models and gives conditions when the current stability analysis results do hold and when they fail in their full extent for the underlying NL system. In Section \ref{sec:examplefull}, the results of Section \ref{sec:lyap} are demonstrated on an example system. Finally, in Section \ref{sec:conclusion}, conclusions on the provided results are given.}

\textit{Notation:}  
The notation $A \succ 0$ $(A \succeq 0)$ indicates that $A$ is positive \mbox{(semi-)definite}, while $A \prec 0$ $(A \preceq 0)$ indicates negative (semi-)definite. The set of $n\times n$ symmetric matrices is denoted by $\mathbb{S}^n$. {$\norm{\cdot}$ is an arbitrary norm over $\mathbb{R}^n$.} The notation $\col(x_1,\dots,x_n)$, denotes the column vector $\begin{bmatrix}	x_1^\top & \cdots & x_n^\top \end{bmatrix}^\top$.

\vspace{-.5em}
\section{Problem Setting}\label{sec:problem}
\subsection{LPV embedding of NL systems}
Consider the NL dynamical system described by
\begin{equation}\label{eq:nlsys}
\begin{aligned}
\dot{x}(t)&=f(x(t),w(t),u(t));\\
z(t)&=h_\textrm{z}(x(t),w(t),u(t));\\
y(t)&=h_\textrm{y}(x(t),w(t));
\end{aligned} 
\end{equation}
where $x(t)\in \mathbb{R}^{n_\mr{x}}$ is the state, $u(t)\in \mathbb{R}^{n_\mr{u}}$ is the control input, $y(t)\in \mathbb{R}^{n_\mr{y}}$ is the measured output, $z(t)\in \mathbb{R}^{n_\mr{z}}$ is the performance variable, $w(t) \in \mathbb{R}^{n_\mr{w}}$ is the disturbance and $t \in \mathbb{R}$ is time. The functions $f:\mathbb{R}^{n_\mr{x}}\times \mathbb{R}^{n_\mr{w}}\times \mathbb{R}^{n_\mr{u}} \rightarrow \mathbb{R}^{n_\mr{x}}$, $h_\mr{z}:\mathbb{R}^{n_\mr{x}}\times \mathbb{R}^{n_\mr{w}}\times \mathbb{R}^{n_\mr{u}} \rightarrow \mathbb{R}^{n_\mr{z}}$ and $h_\mr{y}:\mathbb{R}^{n_\mr{x}}\times \mathbb{R}^{n_\mr{w}} \rightarrow \mathbb{R}^{n_\mr{y}}$, are assumed to be Lipschitz continuous.

This study focuses on analyzing the case when an NL controller for \eqref{eq:nlsys} is designed/analyzed using the LPV framework such that stability and performance guarantees are ensured with respect to the closed-loop behavior $w\rightarrow z$ in order to achieve asymptotic output tracking and disturbance rejection. An LPV model is commonly described by 
\begin{align}\label{eq:lpvsys}
\dot{x}(t) &= A(\rho(t))x(t) + B_\mr{w}(\rho(t))w(t)+ B_\mr{u}(\rho(t))u(t);\notag \\
z(t) &= C_\mr{z}(\rho(t))x(t) + D_\mr{zw}(\rho(t))w(t)+ D_\mr{zu}(\rho(t))u(t);\notag \\
y(t) &= C_\mr{y}(\rho(t))x(t) + D_\mr{yw}(\rho(t))w(t);
\end{align}
where $\rho(t)\in\mathbb{R}^{n_\rho}$ is the scheduling-variable. As explained in Section~\ref{sec:intro}, following the concept of differential inclusions, an NL system \eqref{eq:nlsys}, {under appropriate assumptions,} can be represented in terms of an ``equivalent'' LPV model \eqref{eq:lpvsys}, by appropriately introducing $\rho$, which is a function of the state, output and input variables (or their subset) through a scheduling map $\mu: \mathbb{R}^{n_\mr{x}}\times \mathbb{R}^{n_\mr{w}}\times\mathbb{R}^{n_\mr{u}}\times \mathbb{R}^{n_\mr{y}}\rightarrow \mathbb{R}^{n_\rho}$, such that $\rho(t) = \mu(x(t),w(t),u(t),y(t))$.
Moreover, it is assumed that $\rho(t)$ is confined to a compact convex set $\mathbb{P}\subset \mathbb{R}^{n_\rho}$, such that $\rho(t)\in \mathbb{P}$. Hence, the embedding of the solution set of \eqref{eq:nlsys} is constructed on the compact sets $\mathbb{X}$, $\mathbb{W}$, $\mathbb{U}$ and $\mathbb{Y}$, where $x(t)\in \mathbb{X}$, $w(t) \in \mathbb{W}$, $u(t)\in \mathbb{U}$ and $y(t) \in \mathbb{Y}$, such that $\mu(\mathbb{X},\mathbb{W},\mathbb{U},\mathbb{Y}) \subseteq \mathbb{P}$.
 For simplicity, our analysis will consider consider two cases: when $\mu : \mathbb{X} \rightarrow \mathbb{P}$, i.e. $\rho(t) = \mu(x(t))$, which we will call the ``dependent'' scheduling-variable case and when $\mu : \mathbb{W} \rightarrow \mathbb{P}$, i.e. $\rho(t) = \mu(w(t))$, and hence $\rho$ depends on an external independent signals, which we will call the ``independent'' case.
  See \cite{Kwiatkowski2006,Toth2010,Abbas2014} for several procedures to embed the dynamics of NL systems in an LPV model. 
Based on the LPV model \eqref{eq:lpvsys}, which serves as a proxy description of the NL system \eqref{eq:nlsys}, a controller is synthesized such that the interconnection of controller and {the} LPV model is asymptotically stable and the desired performance criteria on the performance channel from $w\rightarrow z$ are ensured for all $\rho(t)\in \mathbb{P}$. Powerful methods and performance shaping techniques exist to synthesize LPV controllers via convex optimization, see \cite{Packard1993,Apkarian1995,Wu1995,Scherer2001}. In this case, a dynamic output feedback controller is considered of the form
\begin{equation}\label{eq:lpvcontr}
\begin{aligned}
\dot{x}_\mr{c}(t)&=A_\mr{c}(\rho(t))x_\mr{c} (t)+B_\mr{c}(\rho(t)) u_\mr{c}(t);\\
y_\mr{c}(t)&=C_\mr{c}(\rho(t))x_\mr{c} (t)+D_\mr{c}(\rho(t)) u_\mr{c}(t);
\end{aligned}
\end{equation}
where $x_{\mr{c}}(t)\in\mathbb{R}^{n_{\mr{x_c}}}$ is the state, $u_{\mr{c}}(t)\in\mathbb{R}^{n_{\mr{u_c}}}$ the input and $y_{\mr{c}}(t)\in\mathbb{R}^{n_{\mr{y_c}}}$ the output of the controller, respectively.
The controlled LPV system, defined by interconnecting the LPV controller \eqref{eq:lpvcontr} with \eqref{eq:lpvsys}, by taking $u(t)\equiv y_\mr{c}(t)$ and $u_\mr{c}(t) \equiv y(t)$, admits the following description
\begin{equation}\label{eq:lpvclp}
\begin{aligned}
\dot{\xi}(t)&=\m{A}(\rho(t)) \xi(t)+\m{B}(\rho(t))w(t);\\
z(t)&=\m{C}(\rho(t))\xi(t)+\m{D}(\rho(t)) w(t);
\end{aligned}
\end{equation}
with state $\xi(t)= \col\left(x(t),x_\mr{c} (t)\right) \in {\Xi }\subseteq \mathbb{R}^{n_\xi}$\PK{.}

The closed-loop system \eqref{eq:lpvclp} is in fact a proxy description of the NL closed-loop system given by the interconnection of the NL system \eqref{eq:nlsys} and LPV controller \eqref{eq:lpvcontr} with the accompanying scheduling map $\mu$. The NL closed-loop system is then obtained by substituting the scheduling map back in \eqref{eq:lpvclp} resulting, for the dependent case, in
\begin{equation}\label{eq:nlclp}
\begin{aligned}
\dot{\xi}(t)&=\m{A}(\mu(x(t)) \xi(t)+\m{B}(\mu(x(t))w(t);\\
z(t)&=\m{C}(\mu(x(t))\xi(t)+\m{D}(\mu(x(t))) w(t).
\end{aligned}
\end{equation}

Through the LPV embedding \eqref{eq:lpvclp} of the NL system \eqref{eq:nlclp}, convex tools can be used to synthesize the (LPV) controller and guarantee stability of the NL closed-loop system.

The LPV framework had tremendous success in aerospace engineering and the automotive industry with several impactful applications, see \cite{Mohammadpour2012}. However, it has been observed recently in \cite{Scorletti2015} {that, when applying this procedure to guarantee reference tracking and disturbance rejection, the} resulting closed-loop system can exhibit oscillations around the state equilibrium point in the presence of bounded input disturbances $w$ even though asymptotic stability is guaranteed during synthesis, $\rho(t)=\mu(x(t))$ resides in the set $\mathbb{P}${,} and integral action is present. 
{Hence,} the question arises why the LPV controller is unable to achieve asymptotic reference tracking and disturbance rejection when interconnected to the NL system. We will first demonstrate this phenomenon by means of a simple example.\vspace{-.3em}
\subsection{When the implied stability guarantee fails}\label{sec:example}\vspace{-.3em}
\subsubsection{Control scenario}
\begin{figure}
	\vspace{.2em}
	\centering
	\includegraphics[scale=1]{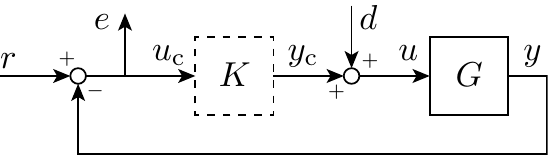}
	\caption{Closed-loop interconnection of plant $G$, \eqref{eq:NL}, and controller $K$, \eqref{eq:controller}.\vspace{-.7em}}
	\label{fig:clpic}
\vspace{-1.5em}
\end{figure}

Consider the following NL system
	\begin{equation}\label{eq:NL}
\begin{aligned}
	\dot{x}(t) &= -x(t)-x^3(t)+u(t);\\
	y(t)&=x(t);
\end{aligned}
\end{equation}
with $x(t),\,u(t),\,y(t)\in\mathbb{R}$. We aim to design an LPV controller in order to achieve reference tracking and disturbance rejection for this system. A possible closed-loop interconnection to achieve this objective is depicted in Fig. \ref{fig:clpic}. We define the generalized disturbance $w = \col(r,d)$, where $r$ is the reference and $d$ the input disturbance, and the generalized performance $z=e$, where $e=r-y$ is the tracking error.

In order to design an LPV controller for our plant and analyze the corresponding closed-loop interconnection, the plant \eqref{eq:NL} is embedded in an LPV model. A possible LPV embedding for \eqref{eq:NL} is
\begin{equation}\label{eq:lpv}
\begin{aligned}
	\dot{x}(t) &= -(1+\rho(t))x(t)+u(t);\\
	y(t) &= x(t);
\end{aligned}
\end{equation}
where $\rho(t) \in \mathbb{P}$ is the scheduling-variable and (we assume) $\mathbb{P} = [0,9]$. The corresponding scheduling map $\mu$ is given by $\rho(t) = \mu(x(t)) = x^2(t) = y^2(t)$.

In order to achieve our control objectives, we consider a PI-like LPV controller, given by 
\begin{equation}\label{eq:controller}
\begin{aligned}
	\dot{x}_\mr{c}(t) &= u_\mr{c}(t);\\
	y_\mr{c}(t) &= (k_{11}+k_{12}\rho(t))x_c(t)+k_{21}u_\mr{c}(t);
\end{aligned}
\end{equation}
where $k_{11}$, $k_{12}$, $k_{21} \in\mathbb{R}$ are parameters of the controller with $x_\mr{c}(t),\, u_\mr{c}(t),\, y_\mr{c}(t)\in\mathbb{R}$. For the (numerical) analysis to follow, the controller parameters are assumed to have the values: $k_{11} = 5$, $k_{12} = 2$ and $k_{21} = 1$. These controller parameter{s} were chosen to demonstrate the stability issues.

The interconnection of \eqref{eq:lpv} and \eqref{eq:controller}, as depicted in Fig. \ref{fig:clpic}, results in an LPV model of the form \eqref{eq:lpvclp} given by
\begin{align}\label{eq:clplpv}
		\dot{x}(t) &= -(1+k_{21}+\rho(t))x(t)+(k_{11}+k_{12}\rho(t))x_\mr{c}(t)\notag\\&\mathrel{\phantom{=}}+k_{21}r(t)+d(t);\notag\\
		\dot{x}_\mr{c}(t) &= -x(t)+r(t);\\
		e(t) &= -x(t)+{r(t)}.\notag
\end{align}
By substituting the scheduling map into \eqref{eq:clplpv}, the corresponding NL closed-loop interconnection is 
\begin{align}\label{eq:clp}
		\dot{x}(t) &= -(1+k_{21})x(t)-x^3(t)+(k_{11}+k_{12}x^2(t))x_\mr{c}(t)\notag\\&\mathrel{\phantom{=}}+k_{21}r(t)+d(t);\notag\\
		\dot{x}_\mr{c}(t) &= -x(t)+r(t);\\
		e(t) &= -x(t)+r(t);\notag
\end{align}
which is a model of the form \eqref{eq:nlclp}.

\subsubsection{$\ltwo$-gain analysis via the LPV concept}
The LPV framework allows for the calculation of an upper bound on the $\ltwo$-gain of \eqref{eq:clp}, by considering \eqref{eq:clplpv} and assuming $\rho \in \mathbb{P}$. Before computing the \ltwo-gain, we connect weighting filters to the inputs and output of the interconnection \eqref{eq:clplpv} in order to incorporate the desired performance specification into our test. To $r$, we connect the weighting filter $W_\mr{r} = 1.5$ (expected magnitude of the reference), to $d$ we connect $W_\mr{d} = 8$ (expected magnitude of the disturbance) and to $e$ we connect $W_\mr{e} = \frac{0.14(s+1)}{s+1\cdot 10^{-7}}$ (sensitivity shaping for integral action and 20\% max overshoot).

\begin{figure}
	\centering
	\includegraphics[scale=1]{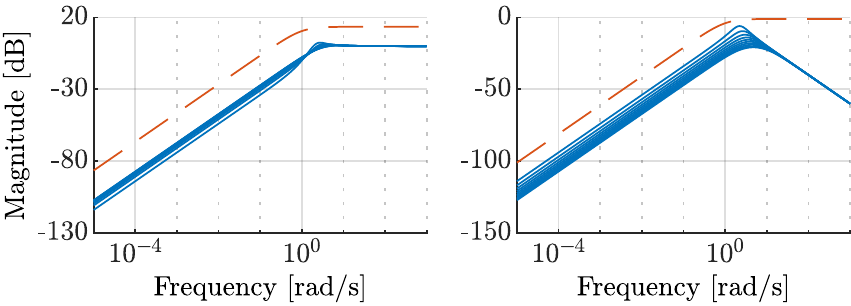}
	\caption{Sensitivity (left, \legendline{mblue}) and process sensitivity (right, \legendline{mblue}) bode magnitude plot for frozen values of the scheduling-variable, including respective inverse weighting filters (\legendline{morange,dashed}).}
	\label{fig:bode}
	\vspace{-1.7em}
\end{figure}

Computing the $\ltwo$-gain of \eqref{eq:clplpv} with the weighting filters connected using the LPVTools Toolbox in MATLAB \cite{Hjartarson2015} results in a $\ltwo$-gain of 0.98. Hence, we can conclude, as long as $\mu(x(t))\in \mathbb{P}$, that \eqref{eq:clp} is $\ltwo$-gain stable and should adhere to the performance specifications defined by the weighting filters. In order to get a sense of the performance of the closed-loop interconnection, the Bode magnitude plot of the sensitivity (i.e. from $r$ to $e$) and process sensitivity (i.e. from $d$ to $e$) against frequency for a number of frozen{\footnote{Constant fixed trajectory of the scheduling-variable, i.e. $\rho(t) \equiv \rho \in \mathbb{P}$. Under such scheduling trajectory \eqref{eq:lpvsys} corresponds to an LTI system, for which a frequency response can be computed.}} values of the scheduling-variable (in $\mathbb{P}$) is given in Fig. \ref{fig:bode}. 

Based on the discussed embedding principle \PK{and as both the sensitivity and process sensitivity for frozen values of the scheduling-variable have magnitudes of zero for a frequency of zero, it would be reasonable to assume, from an LTI analysis point of view,} that for constant reference and disturbance signals (for which still holds that $\rho \in {\mathbb{P}}$) the (interconnected) system \eqref{eq:clp} has zero steady-state error. However, as will be shown next, this is not the case.

\subsubsection{NL time-domain analysis}
\begin{figure}
	\centering
	\includegraphics[scale=1]{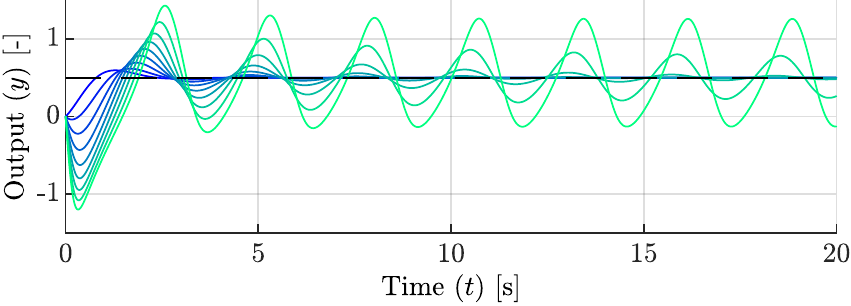}
	\caption{Time responses $y$ of the closed-loop interconnection \eqref{eq:clp} along with the reference $r$ (\legendline{black,dashed}) for constant disturbances $d$ ranging from 0 (\legendline{blue}) to -8 (\legendline{mgreenblue}).	\vspace{-2em}}
	\label{fig:time_response}
\end{figure}

Simulating \eqref{eq:clp} for a constant reference {$r(t)\equiv 0.5$} and various constant disturbances {$d$} results in the time responses displayed in Fig. \ref{fig:time_response}. From Fig. \ref{fig:time_response} it is apparent that for input disturbances closer to zero the output goes to the reference (and we have zero steady-state error). However, applying $d(t)=-7$ or $d(t)=-8$ results in trajectories that converge to orbit-stable limit cycles around the target reference trajectory. Note that based on the trajectories of $y(t)$ in Fig. \ref{fig:time_response}, the corresponding scheduling trajectory stays {within} $\mathbb{P} = [0,9]$ as $y(t) \in [-3,3]$.
Hence, while we adhere to the weighting filters and the scheduling-variable $\rho$ stays within the specified set $\mathbb{P}$, we showed that we do not obtain the expected \PK{desired} behavior.
Thus, by means of this simple example, we have demonstrated that the current $\ltwo$-gain/asymptotic stability and performance analysis through the LPV framework is unfortunately inadequate to imply the tracking and rejection properties for NL systems in general. Next, it will be analyzed why this is the case.

\section{Stability Analysis}
\label{sec:lyap}\subsection{LPV stability analysis}\label{sec:lyaplpv}
Consider the LPV system given by \PK{\eqref{eq:lpvclp}}
with $\rho(t)\in \mathbb{P}$. Define $(\bar{\xi},\bar{w},\bar{\rho})$ to be an equilibrium point of \eqref{eq:lpvclp}, for a given $\bar{w} \in \bar{W}$ and  $\bar{\rho} \in \mathbb{P}$ such that\vspace{-.5em}
\begin{equation}\label{eq:EqP}\vspace{-.2em}
0={\begin{bmatrix}
	\m{A}(\bar{\rho})) & \m{B}(\bar{\rho}))
	\end{bmatrix}}
\begin{bmatrix}
\bar{\xi}\\ \bar{w}
\end{bmatrix}, 
\end{equation}                                  
holds. To simplify the analysis, we assume that for a given $\bar{w}$ and $\bar{\rho}$ there is a unique $\bar{\xi}$ that satisfies \eqref{eq:EqP}. 

\subsubsection{Stability of the origin}\label{sec:lpvstabanalysis}
As seen from \eqref{eq:EqP}, the origin is an equilibrium point of the LPV system \eqref{eq:lpvclp}, by which we mean that $(\bar{\xi},\bar{w},\bar{\rho})=(0,0,{\bar\rho})$ is an equilibrium point for all ${\bar\rho}\in \mathbb{P}$.
Using \PK{standard Lyapunov stability theory}{,} it can be shown, see \cite{Becker1993}, that the origin of the LPV system is quadratically stable if there exists a quadratic Lyapunov function\vspace{-1em}
\begin{equation}\label{eq:lyapfuncor}
	V(\xi) = \xi^\top \m{X}\xi
\end{equation}
with $\m{X}\in \mathbb{S}^{n_\xi}$, where $\m{X}\succ 0$, such that
\begin{equation}\label{eq:vdotorign}
\dot{V}(\xi) = \xi^\top (\m{A}(\rho)^\top \m{X} + \m{X}\m{A}(\rho))\xi \leq 0,
\end{equation}
along all trajectories $\xi(t)$ and $\rho(t)$ of the unperturbed system \eqref{eq:lpvclp}, i.e. \eqref{eq:lpvclp} with $w(t)=0$. If \eqref{eq:vdotorign} is only zero when $\xi=0$, then \eqref{eq:vdotorign} implies asymptotic  stability of \eqref{eq:lpvclp}. Equivalently:
\begin{equation} \label{eq:lyp1}
\m{A}(\rho)^\top\m{X}+\m{X}\m{A}(\rho)\prec 0, \quad \forall\, \rho \in \mathbb{P}.
\end{equation}
As $\mathbb{P}$ is considered {to be} a compact convex set, the infinite dimensional LMI problem \eqref{eq:lyp1} can be reduced to a finite dimensional problem and \PK{assuming $\m{A}(\rho)$ is a convex function} can be solved efficiently using various semidefinite programming solvers, e.g. \cite{Toh1999}.

\subsubsection{Stability of non-zero equilibrium points}\label{sec:lpvnonzeroan}
For quadratic stability of equilibrium points other than the origin consider the Quadratic Lyapunov Function \begin{equation}
{V}_{\bar{\xi}}(\xi) =(\xi-\bar{\xi})^\top \m{X}(\xi-\bar{\xi}){,}\label{eq:quad_lyap_func}
\end{equation} 
with $\m{X}\in \mathbb{S}^{n_\xi}$, where $\m{X}\succ 0$.
Thus, based on \eqref{eq:lpvclp} with $w(t) = \bar{w}$, we obtain that
\begin{align*}
\dot{V}_{\bar{\xi}}(\xi) &= \dot{\xi}^\top \m{X} (\xi-\bar{\xi})+(\xi-\bar{\xi})^\top \m{X} \dot{\xi},\notag\\
&= \left(\m{A}(\rho)\xi+\m{B}(\rho)\bar{w}\right)^\top \m{X}(\xi-\bar{\xi})\notag\\
&\phantom{=}+(\xi-\bar{\xi})^\top \m{X}\left(\m{A}(\rho)\xi+\m{B}(\rho)\bar{w}\right),\notag
\end{align*}
\begin{align}\label{eq:vdotindep}
&= 2(\xi-\bar{\xi})^\top \m{X}\m{A}(\rho)\xi + 2(\xi-\bar{\xi})^\top \m{X}\m{B}(\rho)\bar{w},\notag\\
&= 2(\xi-\bar{\xi})^\top \m{X}\m{A}(\rho)(\xi-\bar{\xi})+2(\xi-\bar{\xi})^\top \m{X}\m{A}(\rho)\bar{\xi}\notag \\
&\phantom{=}+ 2(\xi-\bar{\xi})^\top \m{X}\m{B}(\rho)\bar{w},\notag\\
&= (\xi-\bar{\xi})^\top \underbrace{(\m{A}(\rho)^\top \m{X} + \m{X}\m{A}(\rho))}_{Q(\rho)}(\xi-\bar{\xi})\notag\\
&\phantom{=}+2(\xi-\bar{\xi})^\top \m{X}\underbrace{(\m{A}(\rho)\bar{\xi}+\m{B}(\rho)\bar{w})}_{{Z}(\rho)}.
\end{align}
Hence, to realize the equilibrium point in \eqref{eq:EqP} $\lim\limits_{t\rightarrow\infty}\rho(t)=\bar{\rho}\in \mathbb{P}$, which in turn implies $\lim\limits_{t\rightarrow\infty}Z(\rho(t)) = 0$ (by \eqref{eq:EqP}). Note that in case $\lim\limits_{t\rightarrow\infty}\rho(t)$ does not exist, then it is not possible to prove asymptotic stability of \eqref{eq:lpvclp} w.r.t. an equilibrium point and only relaxed notions such as boundedness may hold under restricted variations of $\rho(t)$.
 Consequently, continuing from \eqref{eq:vdotindep} using $\lim\limits_{t\rightarrow\infty}Z(\rho(t)) = 0$, we obtain
\begin{equation}\label{eq:vdotindepr}
	\dot{V}_{\bar{\xi}}(\xi)=(\xi-\bar{\xi})^\top Q(\rho)(\xi-\bar{\xi}).
\end{equation}
Therefore, if $\lim\limits_{t\rightarrow\infty}\rho(t)=\bar{\rho}$ {and} if there exists $\m{X}\succ 0$ such that
 \begin{equation}\label{eq:lpvquad}
	Q(\rho)=\m{A}(\rho)^\top \m{X} + \m{X}\m{A}(\rho) \prec 0,\quad \forall\,\rho\in \mathbb{P},
\end{equation}
the equilibrium $(\bar{\xi},\bar{w},\bar{\rho})$ is asymptotically stable.

This result is equivalent with \eqref{eq:lyp1}, hence, this means asymptotic convergence to any equilibrium point $(\bar{\xi},\bar{w},\bar{\rho})$, satisfying \eqref{eq:EqP} if $\lim\limits_{t\rightarrow\infty}\rho(t)=\bar{\rho}\in \mathbb{P}$. Due to this property, which similarly holds in the LTI case, LPV stability analysis and performance analysis is accomplished with respect to the origin only (using \eqref{eq:lyapfuncor} and \eqref{eq:vdotorign} as a Lyapunov condition or \eqref{eq:lyapfuncor} as a storage function) as it implies the same guarantees for any other equilibrium point.

\subsection{{NL stability guarantees under state dependent scheduling}}
{As shown in the last section, the standard LPV stability analysis is based on the  assumption that $\rho$} is an exogenous variable, independent of the system dynamics. {In the literature, through the concept of embedding, it is argued that till $\rho(t)\in\mathbb{P}$, stability conclusions made via \eqref{eq:lpvquad} do imply stability of the embedded NL system even in case $\rho=\mu(x)$, similarly as in differential inclusions. As we will show, there is one weak link in the chain of reasoning when $\bar{\xi}\neq0$.}


Assuming that {$\rho$} is only dependent on the state $x$ of the NL system, i.e. $\rho(t) = \mu(x(t))$, results in the closed-loop NL system \PK{\eqref{eq:nlclp}}.
{Furthermore, consider that stability analysis via the LPV form \eqref{eq:lpvclp} of \eqref{eq:nlclp} has been already conducted resulting in a $\m{X}\succ 0$ and $Q(\rho)\prec 0$,  $\forall\,\rho\in \mathbb{P}$. Now, we will investigate what can be concluded based on these relations for the stability of the NL system.

\subsubsection{Stability of the origin}
Performing the stability analysis {for the origin of \eqref{eq:nlclp}, now given by $(\bar{x},\bar{w}) = (0,0)$, that is equivalent with $(0,0,\mu(0))$ in terms of \eqref{eq:EqP}, using the quadratic Lyapunov function \eqref{eq:lyapfuncor} gives}
\begin{equation}
	\dot{V}(\xi) = \xi^\top (\m{A}(\mu(x))^\top \m{X} + \m{X}\m{A}(\mu(x)))\xi.
\end{equation}
{Hence, if the following conditions are satisfied:
\begin{itemize}
\item for the LPV embedding \eqref{eq:lpvclp} of \eqref{eq:nlclp}, there exist a $\m{X}\succ 0$ such that \eqref{eq:lyp1} holds; 
\item $\mu(\mathbb{X})\subseteq \mathbb{P}$, with $\mathbb{X}$ including the origin;
\end{itemize}
then} \eqref{eq:nlclp} is asymptotically stable, as \eqref{eq:lyp1} holds for all $x\in\mathbb{X}$. 

Therefore, asymptotic stability of the origin of the LPV embedding implies asymptotic stability of the origin of the corresponding NL system.

\subsubsection{Stability of non-zero equilibrium points}
Now performing the same stability analysis for non-zero equilibrium points of \eqref{eq:nlclp}, now given by $(\bar{x},\bar{w})$, that, in case of $\rho$ dependent on $x$ is equivalent to $(\bar{x},\bar{w},\mu(\bar{x}))$ in terms of \eqref{eq:EqP}, using the quadratic Lyapunov function \eqref{eq:quad_lyap_func} gives}\footnote{Note that $\xi$ and $x$ are related as $\xi = \col (x,x_\mr{c})$.}
\begin{equation}\label{eq:vdotmux}
	\begin{aligned}
		\dot{V}_{\bar{\xi}}(\xi)&= (\xi-\bar{\xi})^\top Q(\mu(x))(\xi-\bar{\xi})+2(\xi-\bar{\xi})^\top \m{X}Z(\mu(x)),
	\end{aligned}
\end{equation}
However, we only know \eqref{eq:lpvquad}, which does not imply negativity of \eqref{eq:vdotmux} along {all trajectories} $(x(t),\mu(x(t))$. This is only implied if $\lim\limits_{t\rightarrow\infty} \rho(t) = \lim\limits_{t\rightarrow\infty}\mu(x(t)) = \mu(\bar{\xi})$, which {is} not imposed by \eqref{eq:lpvquad}. 
Continuing the analysis of \eqref{eq:vdotmux} and taking $\tilde{\xi} = \xi-\bar{\xi}$, \eqref{eq:vdotmux} can be written as
\begin{equation}
	\tilde{\xi}^\top Q(\mu({x}))\tilde{\xi}+2\tilde{\xi}^\top \m{X}Z(\mu({x})),
\end{equation}
which for any fixed $x\in \mathbb{X}$ is a quadratic matrix polynomial. This quadratic form has as its global maximum at
\begin{equation}\label{eq:lyapmax}
	(\m{X}Z(\mu({x})))^\top (-Q(\mu({x})))^{-1}(\m{X}Z(\mu({x}))).
\end{equation}
As we enforce by \eqref{eq:lpvquad} that $Q(\rho)\prec 0$, the maximum of \eqref{eq:lyapmax} will always be non-negative, hence, there will always be parts of the state-space where the Lyapunov function increases. 
Therefore, based on this analysis, no guarantees for (asymptotic) stability of the equilibrium point can be given {in the general case} if we rely on the results of the LPV test constructed Lyapunov function. Thus, based on the LPV test here, there is no guarantee that the corresponding NL system will be asymptotically stable for an arbitrary equilibrium point, but only for the origin, where the asymptotic stability guarantees are ensured for any arbitrary trajectory of {$\rho$ in $\mathbb{P}$}. Note that in case $\rho=\mu(w)$, i.e. in case of an independent scheduling-variable, $\rho(t) = \mu(w(t)) = \mu(\bar{w}) = \bar{\rho}$ does hold in terms of the equilibrium point \eqref{eq:EqP}. This implies for the NL system \eqref{eq:nlclp} represented by \eqref{eq:lpvclp} that $\rho$ is an exogenous variable, e.g. temperature or windspeed, for which holds that $\lim\limits_{t\rightarrow\infty}\rho(t)=\bar{\rho}\in \mathbb{P}$, asymptotic stability is guaranteed for any equilibrium point $(\bar{\xi},\bar{w},\bar{\rho})$ satisfying \eqref{eq:EqP}, if \eqref{eq:lpvquad} holds.

While for the case of dependent scheduling-variables there are no guarantees anymore that the system is asymptotically stable when performing reference tracking and disturbance rejection, it could still be the case that for a subset of equilibrium points, \PK{\eqref{eq:vdotmux}} is strictly negative for a subset of the state-space, hence, as long as the trajectory stays within this subset of the state-space, asymptotically stability can still be guaranteed for this set of equilibrium points.
This requires computing where \eqref{eq:vdotindepr} (or \eqref{eq:lyapmax}) is negative or alternatively finding the roots of \eqref{eq:vdotindepr} (or \eqref{eq:lyapmax}). However, even in the case that $\mu(x)$ is a linear or a polynomial mapping, \eqref{eq:vdotindepr} becomes a multivariable polynomial, for which it is difficult to find the roots, even for small examples. Moreover, despite the loss of asymptotic stability, boundedness, as can be observed in Section \ref{sec:example}, can still hold. However, this does not coincide with the expected outcome of the LPV analysis, nor would be a desired objective in synthesis.

Furthermore, this stability analysis is based on the Lyapunov function constructed in the LPV analysis step. Of course, for a given NL  system this does not mean that with an alternative method one could not find a Lyapunov function that actually show{s} stability. Here, we mainly investigated the limitations of the currently widely used LPV stability concept.

\vspace{-.7em}
\section{Example}\label{sec:examplefull}
Based on the (asymptotic) Lyapunov stability guarantees given in Section \ref{sec:lyap}, we aim to show for the example system \eqref{eq:clp} that equilibrium points exist for which asymptotic stability cannot be guaranteed. 

Using standard LPV \ltwo-gain stability analysis \cite{Apkarian1995}, an convex optimization problem is solved in order to obtain 
\begin{equation}\label{eq:lyapval}
	\m{X} = \begin{bmatrix}
		 0.6240  & -0.6951\\
   -0.6951  &  3.1187
	\end{bmatrix}.
\end{equation}
 of the quadratic Lyapunov function \PK{\eqref{eq:lyapfuncor}}, a \ltwo-gain\footnote{Note, that no weighting filters are considered in this case.} of 1.78 and asymptotic stability of \eqref{eq:clplpv} for $\rho \in \mathbb{P} = [0,\,9]$. 
As described in Section \ref{sec:lyap}, due to $\rho$ being dependent on $x$, this result only implies asymptotic stability of the origin of \eqref{eq:clp}. Next, we are interested for which set of equilibrium points the underlying NL system asymptotic stability can be guaranteed using the Lyapunov function \PK{\eqref{eq:quad_lyap_func}}
where $\m{X}$ is \eqref{eq:lyapval}. Computing the set of equilibrium points of \eqref{eq:clp} results in \vspace{-.5em}
\begin{equation}
	\Gamma = \left\lbrace (\bar{\xi},\bar{w}) \in {\mathbb{R}^{n_\xi} \times  \mathbb{R}^{n_\mathrm{w}}} \;\vert\; \bar{\xi}=\Omega(\bar{w})  \right\rbrace,
\end{equation}
where\footnote{Assuming that $k_{11},k_{12}>0$ or $k_{11},k_{12}<0$.}
\begin{equation}
	\Omega(\bar{w}) = \begin{bmatrix}
		\bar{r} & \frac{\bar{r}^3+\bar{r}-\bar{d}}{k_{12}\bar{r}^2+k_{11}}
	\end{bmatrix}^\top{,}
\end{equation}
with $\bar{w} = \begin{bmatrix}
	\bar{r}&\bar{d}
\end{bmatrix}^\top$. Furthermore, we define the sets 
\begin{equation}
\begin{gathered}
	\bar{W} = \left\lbrace \bar{w} \;\vert \; \exists \,\bar{w} \in \mathbb{R}^{n_\mr{w}}, \, (\bar{\xi},\bar{w})\in \Gamma \right\rbrace,\\
	\bar{\Xi} = \left\lbrace \bar{\xi} \;\vert \; \exists \,\bar{w} \in \mathbb{R}^{n_\mr{w}}, \, (\bar{\xi},\bar{w})\in \Gamma \right\rbrace.
\end{gathered}
\end{equation}
{Due to the assumption of $\mathbb{P}$ being a convex and compact set, we only consider a part of the state-space for the analysis. We consider $\xi \in \Xi \subset \mathbb{R}^{n_\xi}$, with $\Xi = \{ \xi = \begin{bmatrix} x^\top &x_{\mr{c}}^\top\end{bmatrix}^\top  \;\vert \;  \mu(x)\in \mathbb{P} \}$.}

{For each element $\bar{w}\in \bar{W}$, the subset of $\Xi$ is computed where $\dot{V}_{\bar{\xi}=\Omega(\bar{w})}(\xi)<0$}, \PK{i.e.}
\begin{equation}
	\m{S}_{\bar{w}} := \left\lbrace \xi \;\vert\;\xi \in \Xi,\, \dot{V}_{\Omega(\bar{w})}(\xi)<0,\,\bar{w} \in \bar{W} \right\rbrace.
\end{equation}
{When we consider only a subset of possible reference and disturbance values $\hat{W}\subseteq \bar{W}$, } the intersection of the {corresponding $\m{S}_{\bar{w}}$ sets gives}
	\PK{$\hat{\m{S}} = \bigcap\limits_{\bar{w}\in\hat{W}} \m{S}_{\bar{w}}$}.
{Hence, as long as $\xi(t) \in \hat{\m{S}}$ and $\bar{w}\in \hat{W}$, the trajectory is guaranteed to converge towards a corresponding $\bar{\xi}$. By computing the largest invariant set (reachability set) $\m{R} \subseteq \hat{\m{S}}$ over inputs with $\bar{w}\in \hat{W}$, the NL system is asymptotically stable under any initial condition $\xi_0 \in \m{R}$. 
As commented on before, analytically computing $\hat{\m{S}}$ will be difficult, even for this example with only two states and polynomial scheduling map. Hence, the computation is performed by gridding $\hat{W}$ and ${\Xi}$. For this example we consider\footnote{Note, the specific ${\Xi}$ taken here is consistent with the considered set of the scheduling-variable, $\mathbb{P}$, as the scheduling-variable is given by $\rho = \mu(x) =  x^2$, with $\rho \in \mathbb{P} = [0,\,9]$.} $\hat{W} = [-2,\,2]\times [-8,\,8]$ and $\xi \in \Xi = [-3,\,3]\times [-3,\,3]$. Furthermore, in order to get an understanding of the range of disturbances $\bar{w}$ for which the system is still asymptotically stable, several (gridded) subsets of $\hat{W}$ are considered given by \PK{
	$\hat{W}_\alpha = \alpha\hat{W}$}
where $\alpha \in [0,\,1]$.
\PK{The set $\m{R}$ is approximated}
\PK{ by }simulating \eqref{eq:clp} for \PK{a wide range of} inputs \PK{with $w(t)\in \hat{W}_\alpha$}. 

\begin{figure}
	\centering\vspace{.3em}
	\includegraphics[scale=1]{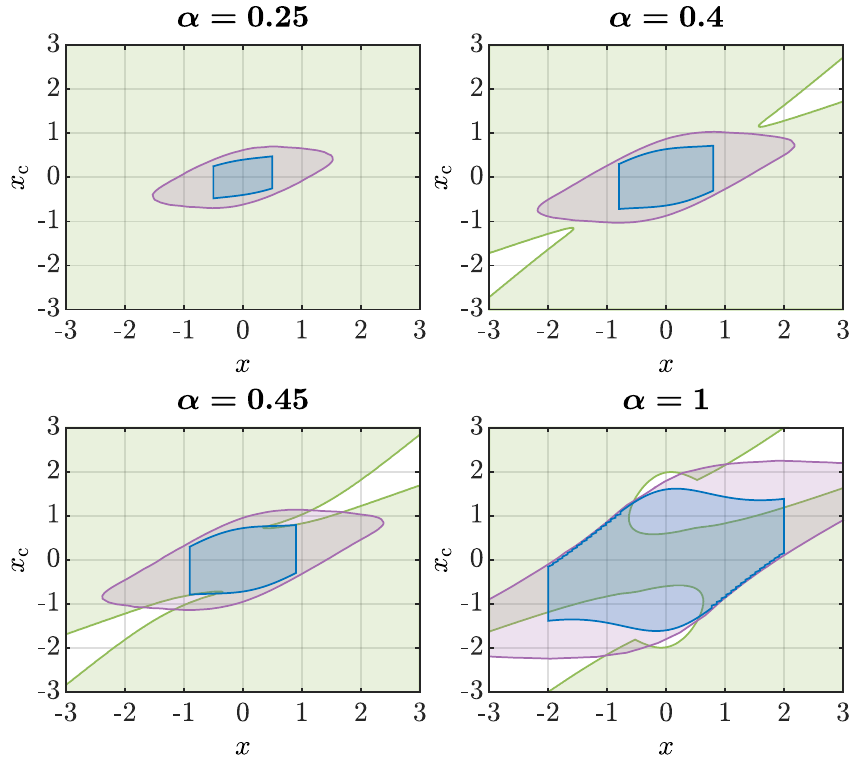}
	\caption{The sets $\hat{\m{S}}$ (\legendsquare{fill=mgreen,fill opacity = 0.2,draw = mgreen}), {$\bar{\Xi}$} (\legendsquare{fill=mblue,fill opacity = 0.2,draw = mblue}), and $\m{R}$ (\legendsquare{fill=mpurple,fill opacity = 0.2,draw = mpurple}), considering $\hat{W}_\alpha$ for different values of $\alpha$.	\vspace{-2em}} 
	\label{fig:setplot}
\end{figure}

In Fig. \ref{fig:setplot} the results are given {for} 
the sets $\hat{\m{S}}$, $\bar{\Xi}$ and $\m{R}$ considering $\hat{W}_\alpha$ for different values of $\alpha$. From the figure it can be observed that \PK{only for approximately $\alpha \leq 0.4$, $\m{R}\subseteq \hat{\m{S}}$, hence, based on this analysis we can only conclude asymptotic stability of the system for $w(t)\in \hat{W}_\alpha$ with $\alpha \leq 0.4$.}
This is in contrast to the $\ltwo$-gain stability guarantee of the LPV model \eqref{eq:clplpv}, for which we guaranteed \PK{asymptotic} stability for all generalized disturbances $w(t)\in\mathbb{R}^2$, \PK{i.e. for $w(t) \in \bar{W}$}. However, as mentioned in Section \ref{sec:lyap}, if the scheduling variable is not independent of the system dynamics there will always exists regions for which asymptotic stability cannot be guaranteed, which, can be observed from Fig. \ref{fig:setplot} for this example.

\section{Conclusion}\label{sec:conclusion}
The LPV framework provides an attractive convex method to check stability of an NL system by considering {its} LPV proxy description. This way of guaranteeing asymptotic stability is also heavily used in synthesizing LPV controller{s} for LPV models of NL plants and using the resulting controller on the NL system. However, the underlying quadratic stability test only ensures asymptotic stability of the origin, which is then argued in the LPV framework to extend to all equilibrium points due to the linearity of the system. In this paper we showed that this fails to hold in the case the scheduling mapping is a function of the state. Hence, for such LPV models of NL systems, asymptotic stability cannot be guaranteed for equilibrium points other than the origin. This means that when applying LPV control methods for an NL system with state dependent scheduling, there are no actual rigorous guarantees when the operation condition changes, e.g. when tracking and rejection is considered. Hence, a different, equilibrium free, stability concept is required that still allows convex synthesis in the LPV framework. The concept of incremental stability is such a notion, which was first used with the LPV framework in \cite{Scorletti2015}, and later extended in \cite{Koelewijn2019a}, to allow for convex controller synthesis ensuring equilibrium independent asymptotic stability.

\addtolength{\textheight}{-12cm}   


\bibliographystyle{IEEEtran}
\bibliography{Ref.bib}

\end{document}